\documentclass[sigconf,9pt]{acmart}
\graphicspath{{./figs/}}

\usepackage{graphicx}
\usepackage{amsmath}
\usepackage{mathtools}
\usepackage{comment}
\usepackage[subrefformat=parens,labelformat=parens]{subfig}
\usepackage{bm}
\usepackage{multirow}
\usepackage{booktabs}
\usepackage{balance}
\usepackage{courier}                                       
\usepackage{cleveref}                                      
\usepackage[mathcal]{eucal}
\usepackage[noend]{algpseudocode}
\algrenewcommand\textproc{\texttt}
\makeatletter\let\float@addtolists\relax\makeatother
\usepackage{algorithm}
\usepackage{siunitx}

\theoremstyle{plain}

\theoremstyle{definition}

\usepackage[skip=1pt]{caption}            
\graphicspath{{./figs/}}

\usepackage{color}
\usepackage{xcolor}
\usepackage{pifont}
\usepackage{makecell}
\usepackage{amsmath}
\usepackage{graphicx}
\usepackage{hyperref}

\definecolor{xiaohan}{cmyk}{0,1.0,1.0,0.45}

\title{APPlace: Co-Optimizing Device Locations and \underline{A}ccess \underline{P}oints for Routing-Aware Analog \underline{Place}ment}

\author{Bingyang Liu}
\affiliation{
  \institution{The University of Texas at Austin}
  \city{Austin}
  \state{TX}
  \country{USA}
}
\email{bingyangliu@utexas.edu}

\author{Xiaohan Gao}
\affiliation{
  \institution{The University of Texas at Austin}
  \city{Austin}
  \state{TX}
  \country{USA}
}
\email{xiaohan.gao@austin.utexas.edu}

\author{Souradip Poddar}
\affiliation{
  \institution{The University of Texas at Austin}
  \city{Austin}
  \state{TX}
  \country{USA}
}
\email{souradippddr1@utexas.edu}

\author{David Z. Pan}
\affiliation{
  \institution{The University of Texas at Austin}
  \city{Austin}
  \state{TX}
  \country{USA}
}
\email{dpan@ece.utexas.edu}

\begin{document}


\copyrightyear{2027}
\acmYear{2027}
\setcopyright{cc}
\setcctype{by}
\acmConference[ASPDAC '27]{32nd Asia and South Pacific Design Automation Conference}{January 25--28, 2027}{Tokyo, Japan}
\acmBooktitle{32nd Asia and South Pacific Design Automation Conference (ASPDAC '27), January 25--28, 2027, Tokyo, Japan}
\acmDOI{10.1145/3840404.3848293}
\acmISBN{979-8-4007-2923-2/2027/01}
\pagestyle{plain}

\begin{abstract}
Analog layout automation remains challenging in Analog/Mixed-Signal circuit design, because parasitics and mismatches introduced during layout strongly affect circuit performance. 
Placement is especially critical among the layout stages, as it constrains the routing solution space and the achievable post-layout performance.
Existing analytical placement methods optimize device coordinates under center-based wirelength proxies and leave access point (AP) selection to routing.
In analog circuits, however, a routing pin is an extended piece of terminal metal rather than a fixed point, and the best AP on it depends on where the other connected devices are placed, so a coordinate-only placement optimizes for connection points that the router will not use.
We present APPlace, a routing-aware analog placement framework that co-optimizes device locations and AP selections during global placement, followed by a legalization stage that preserves the structure established by global placement.
Experimental results on four real-world analog benchmarks show that APPlace achieves better overall post-layout performance than representative baselines at comparable layout area. Ablation studies further show that removing AP co-optimization causes up to a 13~dB CMRR loss among the OTA benchmarks and a 23\% increase in comparator delay.
\end{abstract}



\maketitle



\sloppy
\section{Introduction}
\label{sec:intro}

\begin{figure}[tb]
    \centering
    \includegraphics[width=0.45\textwidth]{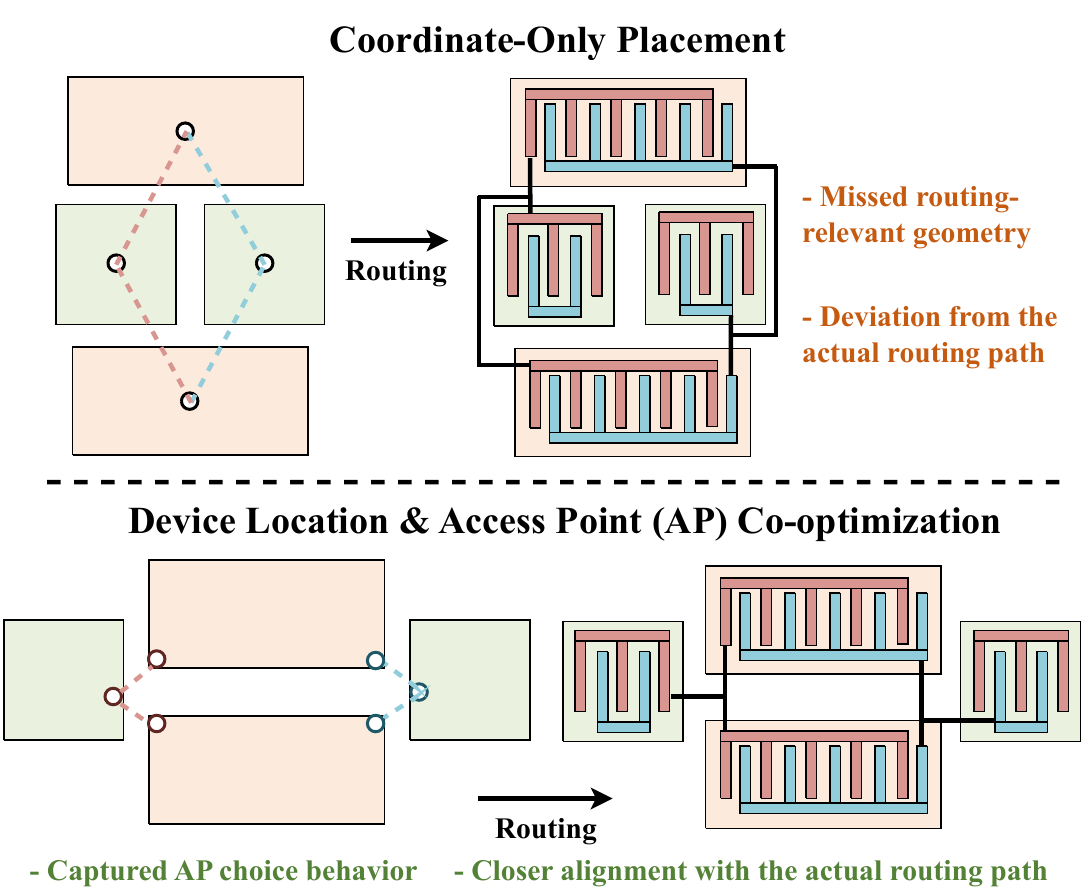}
    \Description{Two rows. Top row, coordinate-only placement: four device blocks are linked by dashed lines between their center points; after routing, the wires attach to the comb-shaped terminal metals at other locations, so the center-based estimate deviates from the actual routing path. Bottom row, device location and access point co-optimization: dashed lines link candidate access points on the terminal edges, and the routed wires follow the chosen access points closely.}
    \caption{Coordinate-only placement can miss routing-relevant geometry. The red and blue circles on the bottom side are candidate APs on the terminal metal. Co-optimizing device locations and APs brings the placement objective closer to the realized routing geometry, which determines parasitics and post-layout performance.
    }
    \label{fig:intro}
    \vspace{-0.3cm}
\end{figure}

In Analog/Mixed-Signal (AMS) circuit design, layout remains a major bottleneck, because parasitics and device mismatches introduced during layout implementation strongly influence the final circuit performance.
In a typical layout flow, placement determines device locations, and routing determines their physical interconnections.
These geometric choices shape the post-layout electrical behavior, yet their effects are difficult to evaluate before the full layout is completed~\cite{budak2023joint}.
As a result, analog layout still relies heavily on manual expertise, which makes effective layout automation both challenging and essential in modern analog design~\cite{xu_magical_2019, chen2021magical, align1, align2, crossley_bag_2013, maji2024toward, chae2025towards, poddar2026_SSCM}.

Among the stages of layout design, placement is particularly critical, as it fixes the geometric positions of devices and thereby restricts the routing solution space and the post-layout performance that routing can ultimately achieve.
Existing analytical placers approximate post-routing quality with wirelength proxies, such as half-perimeter wirelength (HPWL) computed from device centers.
This abstraction is reasonable for digital layouts, where standard cells have fixed pin locations.
In analog layouts, however, a routing pin is an extended piece of terminal metal whose size can be comparable to the routed path itself, and the best point for a net to access it depends on where the other connected devices are placed.
Fixing the access point (AP) at the device center therefore makes placement optimize for a connection point that the router will not actually use, which in turn affects the routed result, the parasitics, and the matching, as illustrated in Fig.~\ref{fig:intro}.

Bringing AP behavior into placement reduces this bias, since the placer then optimizes access geometry close to what the router will realize.
Achieving this in analytical placement faces two challenges.
First, feasible APs are distributed over irregular, non-convex terminal regions, as shown in Fig.~\ref{fig:intro}. This is incompatible with gradient-based global placement (GP), because co-optimizing AP locations with device coordinates in an analytical placer requires a differentiable representation of AP geometry.
Second, the legalization (LG) stage often disrupts the favorable structures formed during GP.
Existing constraint-graph based legalizers~\cite{gao2024joint, gao_interactive_2023, tcad2007koda, xu_magical_2019, xu2019device} target overlap removal under minimum displacement. However, in this way, resolving a few localized conflicts can move the layout globally, which negates the benefit of a routing-aware GP.

To address these challenges, we present APPlace, a routing-aware analog placement framework.
Its GP stage formulates placement as a continuous co-optimization over device coordinates and AP variables, in which the two sets of variables move jointly: where a device sits changes the best APs on its terminals, and the chosen APs in turn pull the devices toward a more routable arrangement.
Its LG stage splits legalization into a reference-anchored global expansion, which creates resolution space while staying tied to the GP result, and a bounded local search, which resolves the remaining sparse conflicts.
This two-stage LG process is designed to minimize the uncontrolled global movement caused by rectifying localized violations.

Our contributions are summarized as follows:
\begin{itemize}
    \item We formulate routing-aware analog global placement as a continuous co-optimization problem over device coordinates and AP variables. A differentiable superellipse AP geometry, an AP-aware mixed wirelength objective, and an overlap-gated adaptive control strategy form this joint optimization.

    \item We propose a structure-preserving legalization algorithm that combines reference-guided expansion with constraint-aware local conflict resolution, which balances global move flexibility and local structural fidelity while eliminating conflicts.

    \item Experimental results on four real-world analog benchmarks show that APPlace achieves stronger overall post-layout performance than MAGICAL~\cite{xu_magical_2019} and Gao et al.~\cite{gao2024joint} under a comparable layout area cost. Ablation study further shows that removing AP co-optimization causes up to a 13~dB CMRR loss among the OTA benchmarks and a 23\% increase in the comparator delay.

    \item APPlace is open-sourced at \url{https://github.com/bingyang1132/applace}.

\end{itemize}

The rest of the paper is organized as follows.
Section~\ref{sec:prelim} reviews related work and the mathematical preliminaries used in our formulation.
Section~\ref{sec:algo} presents the APPlace algorithm.
Section~\ref{sec:exp} reports experimental results.
Section~\ref{sec:conclusion} concludes the paper.

\section{Preliminaries}
\label{sec:prelim}

\subsection{Related Work}

Analog layout automation remains a long-standing challenge due to the strong coupling between versatile constraints and circuit performance~\cite{chen2026anaclara}.
Regarding the placement phase, analog layout automation tools typically adopt one of the mainstream methods:
(1) simulated-annealing (SA) and tree-based placement methods: prior research adapts floorplan techniques, utilizing SA to place devices while satisfying multiple constraints~\cite{iccad2006tam, martins_laygen_2013, ccSymmetry1, align1, ilac, maji2024multiobjective};
(2) template-based methods: template-based tools maintain structural dependency for placement~\cite{crossley_bag_2013, laygo};
(3) analytical placement methods: certain approaches adopt integer linear programming (ILP) for modeling~\cite{xu2017hierarchical, tcad2007koda, iccad2008strasser}, whereas others employ continuous optimization techniques~\cite{xu_magical_2019, zhuHierarchicalAnalogMixedSignal2023, gao2024joint};
(4) machine-learning-based placement methods~\cite{maji2025late}: recent works have explored the use of machine learning models for performance estimation to guide the placement process~\cite{aiplace, smacd2023wang, lin2024rlplace, ispd2026fang}.
A separate line combines placement and routing together within discrete frameworks such as simulated annealing and outputs routed layouts~\cite{cohn1991koan, kubo2002explicit, xiao2010practical, lin2012routability, ou2013simultaneous}. However, they may encounter DRC issues under advanced processes, which are typically resolved by a dedicated router during the placement process described above.

These families differ in solving capability, generality to new circuits, and how directly new objectives can be encoded, where the analytical approach offers comprehensive advantages and have therefore adopted it as the foundation of APPlace.
In analytical placement, device locations and AP variables can be co-optimized as differentiable objectives, for which discrete SA-based search may become costly due to the enlarged variable set considering AP; analytical placement extends to a new circuit without preparing a per-topology template; and performance predictors from the machine-learning line remain complementary, since their predictions can enter the analytical kernel as additional guiding objectives.


Concretely, this analytical formulation casts placement as a continuous optimization that minimizes wirelength with penalty terms enforcing non-overlap and other constraints (e.g., symmetry):

\noindent
\textbf{Wirelength} can be approximated by HPWL. The non-differentiable HPWL representation is usually augmented with a Log-Sum-Exp (LSE) smoothing to have better numerical stability on gradients. The LSE wirelength is defined as follows:

\begin{equation}
W(x) = \sum_{N_j} w_j \gamma \left( \log \sum_{cell_i\in N_j} e^{\frac{x_i}{\gamma}} + \log \sum_{cell_i\in N_j} e^{-\frac{x_i}{\gamma}} \right)
\label{eq:wl_prelim}
\end{equation}

\noindent
\textbf{Non-overlap} is represented as the area of overlapping region:

\begin{equation}
    \begin{aligned}
    O(x) = \sum_{cell_i} \sum_{cell_j} & \max(min(x_i + w_i - x_j, x_j + w_j - x_i), 0)\\ 
    & \cdot \max(min(y_i + h_i - y_j, y_j + h_j - y_i), 0)
    \end{aligned}
\label{eq:ovl_prelim}
\end{equation}

\noindent
\textbf{Symmetry} can be represented with a quadratic term that penalizes symmetry violations. Specifically, the term targets the squared difference in coordinates along the axis of symmetry, as well as the squared differences in the distances from the axis in the perpendicular direction:

\begin{equation}
S(x)= \sum_{cell_i, cell_j} ((x_i + x_j - 2 \cdot x_k^c)^2 + (y_i - y_j)^2) + \sum_{cell_i} (x_i - x_k^c)^2
\label{eq:sym_prelim}
\end{equation}

\subsection{Superellipse Approximation of Rectangle}
\label{subsec:prelim-math}

\begin{figure}[tb]
    \centering
    \includegraphics[width=\linewidth]{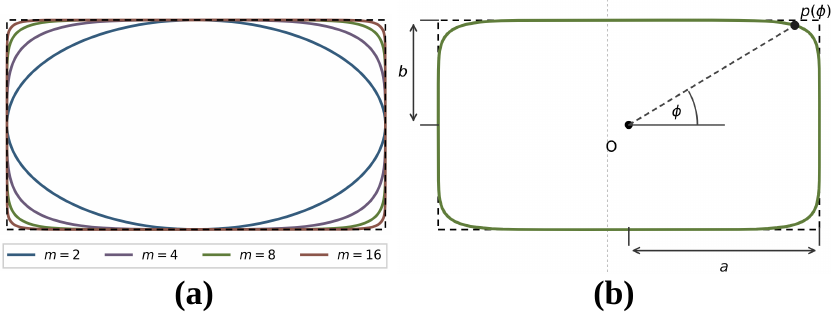}
    \Description{(a) Four nested closed curves inside a dashed rectangle for shape exponents m equal to 2, 4, 8 and 16; m equal to 2 is an ellipse, and larger m gives curves closer to the rectangle. (b) A rounded-rectangle superellipse centered at the origin O with half-width a and half-height b; a dashed ray at angle phi from the horizontal axis meets the boundary at the point p of phi.}
    \caption{
    (a) The transition from ellipse-like to rectangle-like boundaries.
    (b) The geometric meaning of the angular boundary parameter $\phi$.
    }
    \label{fig:superellipse_prelim}
    \vspace{-10pt}
\end{figure}

For horizontal and vertical scales $a>0$ and $b>0$, and a shape exponent $m\ge 2$, define the superellipse region
\begin{equation}
\mathcal{S}_{m}(a,b)
=
\left\{
(x,y)\in\mathbb{R}^{2}
\;\middle|\;
\left|\frac{x}{a}\right|^{m}
+
\left|\frac{y}{b}\right|^{m}
\le 1
\right\}.
\label{eq:superellipse-region}
\end{equation}
When $m=2$, the boundary of $\mathcal{S}_{m}(a,b)$ is an ellipse.
As $m$ increases, the shape becomes progressively more rectangular while remaining smooth for any finite $m$.
Since the underlying $\ell_m$ norm converges to the $\ell_\infty$ norm as $m\to\infty$, $\mathcal{S}_{m}(a,b)$ converges to the rectangle $[-a,a]\times[-b,b]$.

A convenient angular parameterization of the boundary is
\begin{equation}
\begin{aligned}
x(\phi) &= a\,\mathrm{sgn}(\cos\phi)\,|\cos\phi|^{2/m},\\
y(\phi) &= b\,\mathrm{sgn}(\sin\phi)\,|\sin\phi|^{2/m},
\end{aligned}
\qquad \phi\in[-\pi,\pi].
\label{eq:superellipse-param}
\end{equation}
The rectangle $[-a,a]\times[-b,b]$ is a natural geometric approximation of a multi-finger terminal metal bounding box, but its sharp corners are inconvenient for continuous optimization.
With a finite exponent $m$, the superellipse preserves the coarse rectangular support while replacing the corners by a smooth boundary, and larger $m$ yields a tighter rectangle approximation.
In Section~\ref{sec:algo}, we use this family to parameterize AP geometry without optimizing directly over a nonsmooth rectangle.

\section{Algorithm}
\label{sec:algo}

\begin{figure}[tb]
    \centering
    \includegraphics[width=\linewidth]{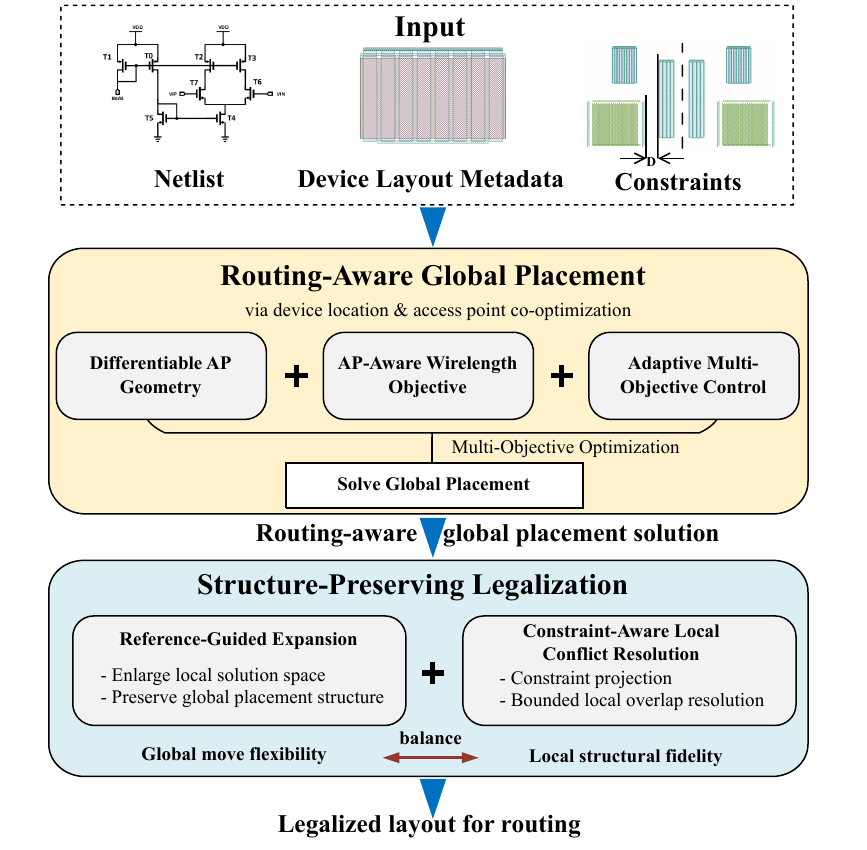}
    \Description{Flowchart. The inputs are a netlist, device layout metadata and placement constraints. They feed a routing-aware global placement stage that combines differentiable access point geometry, an access-point-aware wirelength objective and adaptive multi-objective control to solve global placement. The global placement solution feeds a structure-preserving legalization stage made of reference-guided expansion and constraint-aware local conflict resolution, which balances global move flexibility against local structural fidelity. The output is a legalized layout for routing.}
    \caption{
    APPlace addresses routing-aware analog placement through two stages.
    The first performs routing-aware global placement via device location and access-point co-optimization.
    The second performs structure-preserving legalization to convert the global placement result into a legal layout while preserving the routing-relevant structure.
    }
    \label{fig:overview}
\end{figure}

Figure~\ref{fig:overview} summarizes the workflow of APPlace.
The GP stage formulates placement as a continuous co-optimization over device coordinates and AP variables, and the LG stage converts the GP result into a legal layout that preserves its structure.
The two stages respond to the two challenges raised in Section~\ref{sec:intro}.
For GP, feasible APs are distributed over irregular, non-convex terminal regions, so co-optimizing them with device coordinates by gradients requires a differentiable model of AP geometry (Section~\ref{subsec:algo-gp}).
For LG, resolving localized conflicts must not trigger uncontrolled global movement, so legality has to be reached while the layout stays close to the GP result (Section~\ref{subsec:algo-lg}).
While practical analog placement involves further structural constraints such as alignment and signal flow, we retain the most essential symmetry constraints to keep the formulation centered on the co-optimization of device locations and APs.

\subsection{Routing-Aware Global Placement}
\label{subsec:algo-gp}

\subsubsection{Differentiable AP geometry}
\label{subsubsec:algo-gp-geometry}

We first construct a differentiable representation of AP choice.
For each pin $p$, let $(x_p^{\mathrm{cell}}, y_p^{\mathrm{cell}})$ denote the reference point induced by the layout of its hosting device, and let $(a_p,b_p)$ denote the effective horizontal and vertical scales of the bounding box of the accessible metal region.
Using the superellipse family in Section~\ref{subsec:prelim-math}, we associate each pin with a continuous angular variable $\phi_p$ and define its effective access location as
\begin{equation}
x_p(\phi_p)=x_p^{\mathrm{cell}}+a_p u_p(\phi_p),
\qquad
y_p(\phi_p)=y_p^{\mathrm{cell}}+b_p v_p(\phi_p),
\label{eq:ap-xy}
\end{equation}
where, following the angular parameterization \eqref{eq:superellipse-param},
\begin{equation}
\begin{aligned}
u_p(\phi_p)&=\mathrm{sgn}(\cos\phi_p)\,|\cos\phi_p|^{2/m_p},\\
v_p(\phi_p)&=\mathrm{sgn}(\sin\phi_p)\,|\sin\phi_p|^{2/m_p},
\end{aligned}
\label{eq:ap-uv}
\end{equation}
and $m_p$ is the shape exponent chosen for pin $p$.

Equations~\eqref{eq:ap-xy} and \eqref{eq:ap-uv} convert discrete AP choice into a continuous variable while preserving the coarse rectangular character of terminal metal.
Both $u_p$ and $v_p$ are differentiable in $\phi_p$ almost everywhere, and the device coordinates enter \eqref{eq:ap-xy} through $(x_p^{\mathrm{cell}}, y_p^{\mathrm{cell}})$.
Any objective built on $(x_p, y_p)$ therefore back-propagates to the AP variables and the device coordinates simultaneously.

\subsubsection{AP-aware mixed wirelength}
\label{subsubsec:algo-gp-wirelength}

With the AP geometry in place, we next define a placement objective that gradually incorporates access-aware effects.
The key term is the mixed wirelength
\begin{equation}
W_{\mathrm{mix}}(x,\phi)
=
\bigl(1-\alpha_{\mathrm{ap}}\bigr)W_{\mathrm{base}}(x)
+
\alpha_{\mathrm{ap}} W_{\mathrm{ap}}(x,\phi),
\label{eq:wmix}
\end{equation}
where $W_{\mathrm{base}}$ is the LSE wirelength \eqref{eq:wl_prelim} evaluated on center-based pin coordinates, and $W_{\mathrm{ap}}$ applies the same LSE form to the AP-dependent coordinates in \eqref{eq:ap-xy}:
\begin{equation}
W_{\mathrm{ap}}(x,\phi)
=
\sum_{N_j} w_j \gamma
\left(
\log \sum_{p\in N_j} e^{\frac{x_p(\phi_p)}{\gamma}}
+
\log \sum_{p\in N_j} e^{-\frac{x_p(\phi_p)}{\gamma}}
\right),
\label{eq:wap}
\end{equation}
with the $y$ direction handled analogously.
Through \eqref{eq:ap-xy}, each net in \eqref{eq:wap} pulls devices and APs simultaneously: the gradient with respect to $x_p^{\mathrm{cell}}$ moves the hosting device, and the gradient with respect to $\phi_p$ slides the AP along its terminal boundary.

Equation~\eqref{eq:wmix} controls the transition from the coordinate-level objective to one that increasingly reflects routing-aware geometry.
In early iterations the layout is highly overlapped, and the angular variables $\phi$ have no reliable geometric reference, so a fully AP-aware objective would be overly sensitive.
The mixed objective introduces AP-aware effects gradually.

We therefore update $\alpha_{\mathrm{ap}}\in[0,1]$ with an overlap-gated monotone schedule instead of fixing it as a static hyperparameter.
Let $\bar q_o^{(t)}$ denote the smoothed overlap signal at iteration $t$, and let
$\tau_{\mathrm{gate,lo}}<\tau_{\mathrm{gate,hi}}$
be two gating thresholds.
We define the gate
\begin{equation}
g^{(t)}
=
\operatorname{clamp}
\left(
\frac{\tau_{\mathrm{gate,hi}}-\bar q_o^{(t)}}
{\tau_{\mathrm{gate,hi}}-\tau_{\mathrm{gate,lo}}},
\,0,\,1
\right),
\label{eq:ap-gate}
\end{equation}
and update the AP-mixing coefficient by
\begin{equation}
\alpha_{\mathrm{ap}}^{(t+1)}
=
\min
\left(
\alpha_{\mathrm{ap}}^{(t)}+\eta_{\mathrm{ap}} g^{(t)},
\alpha_{\mathrm{ap}}^{\max}
\right).
\label{eq:ap-schedule}
\end{equation}
Thus, AP-aware effects are activated only after the overlap state has improved sufficiently, and once activated they increase monotonically during GP.

\subsubsection{Global placement objective}
\label{subsubsec:algo-gp-objective}

To improve post-layout quality, analytical placement typically minimizes a wirelength proxy while penalizing overlap and structural violations.
For continuous optimization, we relax these requirements into the weighted objective, similar to~\cite{gao2024joint}, but with the AP-aware wirelength in place:
\begin{equation}
F_{\mathrm{GP}}(x,\phi)
=
W_{\mathrm{mix}}(x,\phi)
+
\lambda_{\mathrm{ovl}} O(x)
+
\lambda_{\mathrm{sym}} S(x),
\label{eq:fgp}
\end{equation}
where $W_{\mathrm{mix}}$ is the AP-aware mixed wirelength objective, $O(x)$ is the overlap penalty, and $S(x)$ is the symmetry penalty.
$W_{\mathrm{mix}}$ is the key new component, as it lets AP-aware geometry enter placement directly, while the overlap and symmetry terms follow standard analytical analog placement formulations~\cite{gao2024joint}.

\subsubsection{Adaptive multi-objective control}
\label{subsubsec:algo-gp-controller}

The GP stage regulates the multipliers in \eqref{eq:fgp} through normalized and smoothed violation signals.
We first define the raw normalized violations
\begin{equation}
q_o^{\mathrm{GP},(t)}=\frac{O_{\mathrm{real}}^{(t)}}{A_{\mathrm{ref}}},
\qquad
q_s^{(t)}=\frac{S^{(t)}}{S_{\mathrm{ref}}},
\label{eq:q-gp}
\end{equation}
and then smooth them over iterations:
\begin{equation}
\bar q_j^{(t)}
=
(1-\rho_j)\bar q_j^{(t-1)}+\rho_j q_j^{(t)},
\qquad
j\in\{o,s\},
\label{eq:qbar}
\end{equation}
where $\rho_j\in(0,1]$ is a smoothing factor.
Smoothing suppresses iteration-level noise, so the controller responds to persistent structural imbalance.

The control response combines the current deviation from the target level with the recent trend:
\begin{equation}
h_j^{(t)}
=
a_j\left(\frac{\bar q_j^{(t)}}{\tau_j}-1\right)
+
b_j
\log\frac{\bar q_j^{(t)}+\varepsilon}{\bar q_j^{(t-1)}+\varepsilon},
\qquad j\in\{o,s\},
\label{eq:hj}
\end{equation}
where $\tau_o$ and $\tau_s$ denote the overlap and symmetry targets.
The first term measures how far a violation is from its target, and the second term distinguishes a violation that is improving from one that is worsening.
The responses are normalized into an update direction
\begin{equation}
(d_o^{(t)},d_s^{(t)})
=
\frac{(h_o^{(t)},h_s^{(t)})}
{\sqrt{(h_o^{(t)})^2+(h_s^{(t)})^2}+\varepsilon},
\label{eq:dj}
\end{equation}
and the multipliers are updated in the log domain:
\begin{equation}
\log\lambda_j^{(t+1)}
=
\operatorname{clip}
\left(
\log\lambda_j^{(t)}+s_\lambda d_j^{(t)}
\right),
\qquad
j\in\{o,s\}.
\label{eq:lambdaupdate}
\end{equation}
The update is multiplicative and therefore responds to relative imbalance, which matters because the relevant multiplier scales differ substantially across instances and optimization phases.

The multipliers are initialized by the gradient-balancing initialization commonly used in analytical placement~\cite{li2019elfplace, gao2024joint}:
\begin{equation}
\lambda_{\mathrm{ovl}}^{(0)}
=
\operatorname{clip}
\left(
\frac{\|\nabla W_{\mathrm{mix}}\|_2}
{\|\nabla O_{\mathrm{real}}\|_2+\varepsilon}
\right),
\qquad
\lambda_{\mathrm{sym}}^{(0)}
=
\operatorname{clip}
\left(
\frac{\|\nabla W_{\mathrm{mix}}\|_2}
{\|\nabla S\|_2+\varepsilon}
\right),
\label{eq:lambda-init}
\end{equation}
so that the major objective terms start on comparable optimization scales.
Algorithm~\ref{alg:gp} summarizes the GP stage.

\begin{algorithm}[t]
\caption{Routing-Aware Global Placement}
\label{alg:gp}
\small
\begin{algorithmic}[1]
\Require Netlist, device geometry, pin location, symmetry constraints
\Ensure Routing-aware global placement solution $(x,\phi)$

\State $(x,\phi) \gets$ initial placement and AP variables; $\alpha_{\mathrm{ap}} \gets 0$
\State $(\lambda_{\mathrm{ovl}},\lambda_{\mathrm{sym}}) \gets$ gradient-balancing initialization \eqref{eq:lambda-init}

\For{$t=1,\dots,T_{\mathrm{GP}}$}
    \State $(x_p,y_p) \gets$ AP-aware pin coordinates \eqref{eq:ap-xy}
    \State $F_{\mathrm{GP}} \gets W_{\mathrm{mix}}(x,\phi) + \lambda_{\mathrm{ovl}} O(x) + \lambda_{\mathrm{sym}} S(x)$ \Comment{\eqref{eq:wmix}, \eqref{eq:fgp}}
    \State $(x,\phi) \gets \mathrm{OptimizerStep}\bigl((x,\phi), \nabla_{(x,\phi)} F_{\mathrm{GP}}\bigr)$
    \State $\alpha_{\mathrm{ap}} \gets \min\bigl(\alpha_{\mathrm{ap}} + \eta_{\mathrm{ap}} g^{(t)},\, \alpha_{\mathrm{ap}}^{\max}\bigr)$ \Comment{\eqref{eq:ap-gate}, \eqref{eq:ap-schedule}}
    \State $(\lambda_{\mathrm{ovl}},\lambda_{\mathrm{sym}}) \gets$ multiplier update \eqref{eq:hj}--\eqref{eq:lambdaupdate}
\EndFor
\State \Return $(x,\phi)$
\end{algorithmic}
\end{algorithm}

\subsection{Structure-Preserving Legalization}
\label{subsec:algo-lg}

\subsubsection{Reference-Guided expansion}
\label{subsubsec:algo-lg-expansion}

Purely global legalization easily destroys the structures established in GP, while purely local resolution often fails in dense overlaps.
Our LG stage therefore first creates resolution space through a reference-guided expansion.
Let $x^{\mathrm{GP}}$ denote the routing-aware global placement result and let $x$ denote the current legalization coordinates.
The expansion stage optimizes
\begin{equation}
F_{\mathrm{LG}}(x)
=
W_{\mathrm{mix}}^{\mathrm{real}}(x)
+
\lambda_{\mathrm{ovl}}^{\mathrm{LG}} O_{\mathrm{infl}}(x)
+
\lambda_{\mathrm{anchor}} H_{\mathrm{ref}}(x)
+
\lambda_{\mathrm{cover}} C_{\mathrm{real}}(x),
\label{eq:flg}
\end{equation}
which enlarges the feasible resolution space while keeping the layout tied to the structure formed in GP.

The first term, $W_{\mathrm{mix}}^{\mathrm{real}}(x)$, keeps the layout from drifting away from a routing-relevant geometry.
Its role here is weaker than in GP, which is to prevent unnecessary distortion while expansion is taking place. The second term, $O_{\mathrm{infl}}(x)$, evaluates the overlap penalty \eqref{eq:ovl_prelim} on inflated geometry.
If the real dimensions of object $i$ are $(w_i,h_i)$, the inflated dimensions are
\begin{equation}
w_i^{\mathrm{infl}} = w_i + 2\alpha_{\mathrm{infl}}\delta_i,
\qquad
h_i^{\mathrm{infl}} = h_i + 2\alpha_{\mathrm{infl}}\delta_i,
\label{eq:inflated}
\end{equation}
where $\delta_i$ is an inflation margin and $\alpha_{\mathrm{infl}}\in[0,1]$ is an expansion schedule, and $O_{\mathrm{infl}}$ substitutes $(w_i^{\mathrm{infl}}, h_i^{\mathrm{infl}})$ for $(w_i,h_i)$ in \eqref{eq:ovl_prelim}.
By treating each object as temporarily larger than its real size,
the optimization creates breathing room for the final hard conflict resolution.

A layout encouraged only to open up can drift too freely and lose the structure established by GP.
To prevent this, we anchor the layout to a dilated reference.
Let $p_i^{\mathrm{GP}}$ be the center of object $i$ in $x^{\mathrm{GP}}$, and let $c$ be the global layout center.
We define
\begin{equation}
p_i^{\mathrm{ref}}
=
c+s\bigl(p_i^{\mathrm{GP}}-c\bigr),
\qquad
s=1+\beta\alpha_{\mathrm{infl}},
\label{eq:ref}
\end{equation}
and use the reference penalty
\begin{equation}
H_{\mathrm{ref}}(x)
=
\sum_i \|p_i-p_i^{\mathrm{ref}}\|_2^2
\label{eq:href}
\end{equation}
to keep the expanded layout close to this dilated configuration.
If anchored directly to $x^{\mathrm{GP}}$, the optimizer would resist the very opening-up behavior that legalization needs.
By expanding the reference outward as $\alpha_{\mathrm{infl}}$ increases, \eqref{eq:ref} allows the layout to move globally while still preserving the coarse topology inherited from GP.

Standard overlap penalties also fail to resolve burying configurations in which a small device is almost fully covered by a larger one, because the raw overlap signal is dominated by scale imbalance.
Simply increasing the global overlap weight disturbs the overall multi-objective balance more than it helps the buried object escape.
We therefore introduce a cover penalty.
For a pair $(i,j)$ with real overlap area $I_{ij}$ and real areas $A_i,A_j$, define
\begin{equation}
r_{ij}^{\mathrm{cover}}=\frac{I_{ij}}{\min(A_i,A_j)}.
\label{eq:rcover}
\end{equation}
Then
\begin{equation}
C_{\mathrm{real}}(x)
=
\sum_{(i,j)\in\mathcal{E}_{\mathrm{ovl}}}
\psi\!\left(r_{ij}^{\mathrm{cover}}\right)
\label{eq:creal}
\end{equation}
penalizes near-full-cover cases and provides a scale-normalized signal for pushing small objects out of pathological overlaps.

\subsubsection{Constraint-aware local conflict resolution}
\label{subsubsec:algo-lg-resolution}

After the reference-guided expansion, most severe conflicts have been relaxed, and the remaining violations are typically sparse and localized.
We resolve them through a local stage that first converts the soft symmetry obtained in GP and expansion into exact geometric symmetry, and then performs bounded local resolution.

The symmetry projection is handled geometrically rather than through another optimization loop.
Self-symmetric objects are projected onto the designated symmetry axis, and symmetric pairs are aligned and placed equidistantly around that axis.
This step closes the remaining gap between a near-symmetric continuous solution and an exactly symmetric legal layout.

After symmetry is made explicit, we resolve the remaining overlaps by searching nearby feasible candidate moves and iteratively applying the best one.
For symmetry-coupled objects, candidate moves are generated and scored jointly to preserve symmetry throughout local resolution.
Each candidate is scored by:
\begin{equation}
J(c)
=
d_{\mathrm{ell}}(c)
+
\lambda_{\mathrm{expand}} e_{\mathrm{expand}}(c),
\label{eq:score}
\end{equation}
where $d_{\mathrm{ell}}(c)$ is an elliptical local movement cost, and $e_{\mathrm{expand}}(c)$ penalizes global bounding-box growth.

The movement term is measured by an elliptical distance,
\begin{equation}
d_{\mathrm{ell}}(c)
=
\sqrt{
\left(\frac{\Delta x(c)}{r_x}\right)^2
+
\left(\frac{\Delta y(c)}{r_y}\right)^2
},
\label{eq:dell}
\end{equation}
where $(\Delta x(c),\Delta y(c))$ is the displacement induced by candidate $c$, and $(r_x,r_y)$ are anisotropic movement scales.
The elliptical form reflects that device and layout aspect ratios are generally not close to one, so horizontal and vertical motion need not have the same geometric cost.

To avoid resolving local conflicts by trivial spreading, we penalize only positive growth of the layout bounding box:
\begin{equation}
e_{\mathrm{expand}}(c)
=
\frac{[\Delta W(c)]_+}{W_{\mathrm{bbox}}}
+
\frac{[\Delta H(c)]_+}{H_{\mathrm{bbox}}}
+
\frac{[\Delta A(c)]_+}{A_{\mathrm{bbox}}},
\label{eq:eexpand}
\end{equation}
where $[u]_+=\max(u,0)$.
This keeps the final resolution local in its action while still accounting for global geometric damage and area cost.

When local conflict resolution cannot eliminate all overlaps in a single pass,
we selectively inflate the expansion margins of the offending devices by a
multiplicative factor $\mu_{\mathrm{retry}}>1$, restore the layout to
$x^{\mathrm{GP}}$, and re-run the full legalization pass.
The enlarged margins widen the search space for the subsequent attempt.
Because the margins grow monotonically, the retry terminates within a bounded
number of rounds $R_{\max}$, and the pass with the fewest residual overlaps is
accepted.
Algorithm~\ref{alg:lg} summarizes the complete legalization stage including
this retry mechanism.

\begin{algorithm}[t]
\caption{Structure-Preserving Legalization}
\label{alg:lg}
\small
\begin{algorithmic}[1]
\Require Routing-aware GP solution $x^{\mathrm{GP}}$, symmetry constraints
\Ensure Legalized layout $x$

\State $x \gets x^{\mathrm{GP}}$; $\delta \gets$ initial inflation margins
\For{round $= 0, 1, \dots, R_{\max}$}
    \For{$t=1,\dots,T_{\mathrm{LG}}$} \Comment{reference-guided expansion}
        \State $(w^{\mathrm{infl}},h^{\mathrm{infl}}) \gets$ inflated geometry \eqref{eq:inflated}; $\ p^{\mathrm{ref}} \gets$ dilated reference \eqref{eq:ref}
        \State $F_{\mathrm{LG}} \gets$ expansion objective \eqref{eq:flg}
        \State $x \gets \mathrm{OptimizerStep}\bigl(x, \nabla_x F_{\mathrm{LG}}\bigr)$
    \EndFor
    \State $x \gets$ exact-symmetry projection
    \While{overlap remains} \Comment{local conflict resolution}
        \State $\mathcal{C} \gets$ feasible candidate moves near conflicting objects
        \State $c^{\ast} \gets \arg\min_{c\in\mathcal{C}} J(c)$ \Comment{\eqref{eq:score}}
        \State $x \gets$ layout after applying $c^{\ast}$
    \EndWhile
    \If{no unresolved overlaps}
        \State \textbf{break}
    \EndIf
    \State $\delta_i \gets \mu_{\mathrm{retry}} \cdot \delta_i$ for each unresolved device $i$
    \State $x \gets x^{\mathrm{GP}}$ \Comment{restore and retry}
\EndFor
\State \Return $x$ from the pass with fewest residual overlaps
\end{algorithmic}
\end{algorithm}

\section{Experiments}
\label{sec:exp}

\subsection{Experimental Setup}
\label{subsec:exp-setup}

\subsubsection{Benchmarks and baselines}
\begin{figure*}[t]
    \centering
    \includegraphics[width=\linewidth]{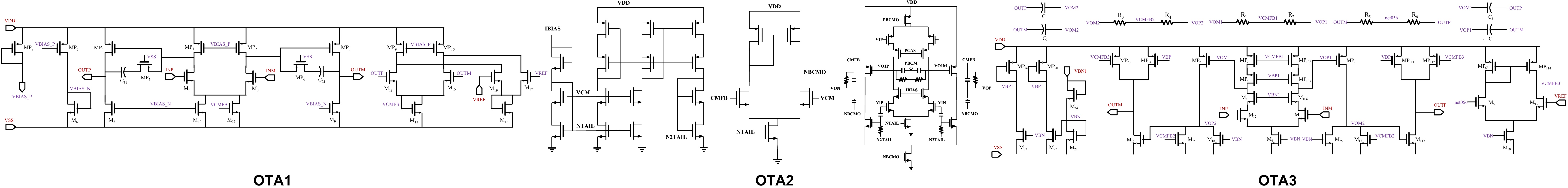}
    \Description{Transistor-level schematics of the three operational transconductance amplifier benchmarks, labeled OTA1, OTA2 and OTA3, drawn side by side.}
    \caption{Schematics of the three OTA benchmarks (OTA1--OTA3).}
    \label{fig:schematic}
    \vspace{-0.3cm}
\end{figure*}

We evaluate our framework on four analog benchmarks implemented in the TSMC N40 technology: three operational transconductance amplifiers (OTA1--OTA3), whose schematics are shown in Fig.~\ref{fig:schematic}, and a clocked dynamic comparator (COMP).
The benchmarks are selected to span a wide range of topological complexity and layout difficulty.
OTA1 is a two-stage amplifier.
OTA2 is a fully differential design with common-mode feedback.
OTA3 is a multi-stage fully differential design with compensation networks and the largest device count.
COMP is a clocked dynamic comparator with a differential input pair and a cross-coupled regenerative latch; its delay and offset are sensitive to layout parasitics.
Table~\ref{tab:bench} lists the size of each benchmark.

\begin{table}[htbp]
\centering
\caption{Benchmark statistics.}
\label{tab:bench}
{\renewcommand{\arraystretch}{0.9}
\footnotesize
\resizebox{0.75\columnwidth}{!}{
\begin{tabular}{lccc}
\toprule
Circuit & \#Devices & \#Nets & \#AP variables \\
\midrule
OTA1 & 26 & 18 & 97  \\
OTA2 & 50 & 35 & 177 \\
OTA3 & 43 & 27 & 165 \\
COMP & 18 & 12 & 69  \\
\bottomrule
\end{tabular}
}}
\vspace{-0.3cm}
\end{table}

\begin{table*}[tb]
\centering
\caption{Post-layout results on OTA1, OTA2, and OTA3.}
\label{tab:ota_all_main}
\resizebox{\textwidth}{!}{
\begin{tabular}{l ccccc ccccc ccccc}
\toprule
& \multicolumn{5}{c}{OTA1} & \multicolumn{5}{c}{OTA2} & \multicolumn{5}{c}{OTA3} \\
\cmidrule(lr){2-6} \cmidrule(lr){7-11} \cmidrule(lr){12-16}
Metric & Schem. & Gao & MAG. & Hyb.$^{\dagger}$ & APPlace & Schem. & Gao & MAG. & Hyb.$^{\dagger}$ & APPlace & Schem. & Gao & MAG. & Hyb.$^{\dagger}$ & APPlace \\
\midrule
Gain (dB) & 49.00 & 49.28 & 49.34 & 48.42 & \textbf{49.65} & 62.04 & 53.15 & 52.41 & 41.55 & \textbf{59.83} & 73.52 & -- & 54.13 & 54.78 & \textbf{59.13} \\
UGB (MHz) & 0.219 & 0.199 & 0.207 & 0.191 & \textbf{0.228} & 2.191 & 1.454 & 1.211 & 1.647 & \textbf{1.734} & 1.924 & -- & 1.132 & \textbf{1.424} & 1.235 \\
CMRR (dB) & 50.85 & 34.06 & 36.19 & 34.64 & \textbf{37.10} & 101.70 & 55.70 & 72.99 & 68.45 & \textbf{88.86} & 84.91 & -- & 54.90 & 67.34 & \textbf{81.26} \\
PM ($^\circ$) & 71.9 & 70.6 & 70.8 & 67.2 & \textbf{71.6} & 75.9 & 59.4 & 62.5 & 73.9 & \textbf{77.1} & 89.7 & -- & \textbf{89.3} & 79.7 & 83.9 \\
WL ($\mu$m) & -- & 2462 & 1327 & 1212 & \textbf{1192} & -- & 3860 & \textbf{3654} & 4039 & 3707 & -- & -- & 4311 & 5116 & \textbf{4040} \\
Area ($\mu$m$^2$) & -- & 12963 & \textbf{2759} & 3719 & 3113 & -- & 7080 & \textbf{5240} & 7625 & 6263 & -- & -- & \textbf{10092} & 20769 & 11864 \\
\bottomrule
\end{tabular}
}
\begin{minipage}{\linewidth}
    \footnotesize
    $^{\dagger}$Hyb.\ (Hybrid): APPlace GP followed by the MAGICAL legalizer. Schem.: pre-layout schematic reference; Gao: Gao et al.~\cite{gao2024joint}; MAG.: MAGICAL~\cite{xu_magical_2019}. Gao et al.\ fails at legalization on OTA3.
\end{minipage}
\vspace{-0.3cm}
\end{table*}

We compare APPlace against two full-flow baselines, {MAGICAL}~\cite{chen_magical_2021, chen2021magical, xu_magical_2019} and {Gao et al.}~\cite{gao2024joint}.
We also evaluate a hybrid flow, denoted {Hybrid}, which keeps the APPlace GP result and replaces only the legalization stage with the MAGICAL legalizer, thereby isolating the contribution of our legalization algorithm.
For post-layout evaluation, we extract parasitics with Calibre and simulate with Cadence Spectre.
The schematic entries in Tables~\ref{tab:ota_all_main} and~\ref{tab:comp_main} report pre-layout performance and serve only as an ideal reference. As for routing, in APPlace and Gao et al., we adopt the routing methodology of SAGERoute~\cite{zhang2023sage,zhang2024sage} for the post-placement stages.
Under this identical routing backend, performance gaps between the two flows can be strictly attributed to placement.
{MAGICAL} is evaluated with its native end-to-end flow, because its internal router relies on its built-in device generator and is incompatible with the standard PDK PCells used by the other flows.
In both routing backends, the router automatically searches and selects APs in the placed layout~\cite{zhang2023sage, zhang2024sage, xu_magical_2019}.

\subsubsection{Evaluation metrics}
Tables~\ref{tab:ota_all_main} and~\ref{tab:comp_main} report post-layout performance, routed wirelength (WL), and layout area.
For the OTAs, the performance metrics are gain, unity-gain bandwidth (UGB), common-mode rejection ratio (CMRR), and phase margin (PM).
For COMP, the metrics are average delay, offset, and energy per cycle.
Fig.~\ref{fig:radar} summarizes the same data as a per-circuit tradeoff view.

\subsection{Overall Post-Layout Quality}
\label{subsec:exp-overall}

\begin{table}[htbp]
\centering
\caption{Post-layout results on COMP.}
\label{tab:comp_main}
\resizebox{\columnwidth}{!}{
\begin{tabular}{lccccc}
\toprule
Flow & Delay (ps) & Offset (mV) & E/cycle (fJ) & WL ($\mu$m) & Area ($\mu$m$^2$) \\
\midrule
Schematic & 55.87 & - & 55.92 & -- & -- \\
MAGICAL & 135.15 & 1.772 & \textbf{90.01} & \textbf{338} & 262 \\
Gao et al. & 146.35 & 4.723 & 99.05 & 439 & 230 \\
Hybrid$^{\dagger}$ & 111.6 & 1.947 & 104.2 & 393 & 230 \\
APPlace & \textbf{109.9} & \textbf{1.682} & 93.90 & 366 & \textbf{228} \\
\bottomrule
\end{tabular}
}
\begin{minipage}{\linewidth}
    \footnotesize
    $^{\dagger}$Hybrid: APPlace GP followed by the MAGICAL legalizer.
\end{minipage}
\vspace{-0.4cm}
\end{table}

\begin{figure}[tb]
    \centering
    \includegraphics[width=0.7\linewidth]{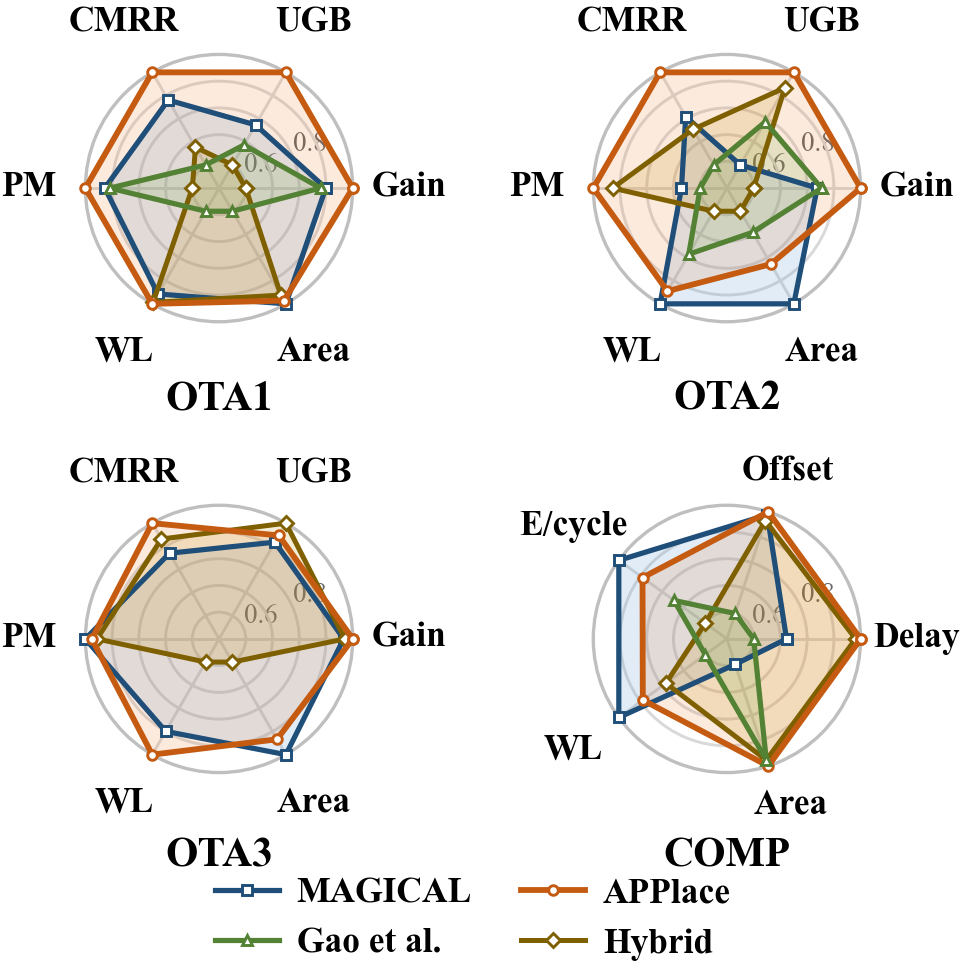}
    \Description{Four radar charts, one each for OTA1, OTA2, OTA3 and COMP, overlaying MAGICAL, Gao et al., Hybrid and APPlace. The OTA charts have axes for gain, UGB, CMRR, PM, WL and area; the COMP chart has axes for delay, offset, energy per cycle, WL and area. Points farther from the center indicate better normalized quality.}
    \caption{
    Post-layout tradeoff comparison on OTA1, OTA2, OTA3 and COMP. Metrics are min-max normalized within each circuit; cost metrics are inverted, with larger radius indicating better quality.
    }
    \label{fig:radar}
\vspace{-0.3cm}
\end{figure}

Tables~\ref{tab:ota_all_main} and \ref{tab:comp_main} report the post-layout results, and Fig.~\ref{fig:radar} visualizes the per-circuit tradeoffs.
{APPlace} achieves the strongest overall post-layout performance on all four benchmarks and avoids the sharp single-metric degradation observed in the baselines.

On OTA1 and OTA2, {APPlace} achieves the best values of all four electrical metrics.
On OTA2, for instance, it improves gain by 7.4~dB and CMRR by 15.9~dB over {MAGICAL} while remaining other metrics comparable.
On OTA3, {Gao et al.} fails at the legalization stage because excessive constraint edges make its linear program infeasible.
Among the completed layouts, {APPlace} provides the best gain, the best CMRR, and the shortest routed wirelength.
On COMP, {APPlace} achieves the best delay, the best offset, and the smallest area among all four flows.
A baseline may peak on an isolated metric, such as the PM of {MAGICAL} on OTA3, yet none provides a comparable overall balance of performance and geometry cost.

The comparison with {Gao et al.} is particularly informative because both flows share the same routing backend.
The consistent gap therefore supports the advantage of joint AP-coordinate optimization over coordinate-only placement followed by downstream AP search.
All APPlace results above are produced with 10.5, 62.4, 25.2, and 7.1 seconds of placement runtime on OTA1, OTA2, OTA3, and COMP, respectively, including both GP and LG, measured on a single thread of an Intel Core i7-12700 (12 cores, 64~GB RAM) running Ubuntu 20.04.
The optimizer is deterministic, so each benchmark produces the same result across runs.
Section~\ref{subsec:exp-ablation} further isolates the contribution of each proposed component.

\subsection{Ablation Studies}
\label{subsec:exp-ablation}

\subsubsection{AP co-optimization}
\label{subsubsec:exp-ablation-ap}
We isolate the effect of AP co-optimization by holding the mixing coefficient $\alpha_{\mathrm{ap}}$ at zero throughout global placement, with the optimizer, the multiplier control, the legalizer, and the router all unchanged.
In this {w/o AP} variant, device coordinates are still optimized, but AP locations are no longer co-optimized with them.
Table~\ref{tab:ap_ablation} compares the two variants on all four benchmarks.

\begin{table}[htbp]
\centering
\caption{Ablation of AP co-optimization.}
\label{tab:ap_ablation}
\resizebox{\columnwidth}{!}{
\begin{tabular}{clcccccc}
\toprule
Design & Variant & Gain (dB) & UGB (MHz) & CMRR (dB) & PM ($^\circ$) & WL ($\mu$m) & Area ($\mu$m$^2$) \\
\midrule
\multirow{2}{*}{OTA1}
& w/o AP  & 45.67 & \textbf{0.234} & 35.70 & 71.6 & 1251 & 3203 \\
& APPlace & \textbf{49.65} & 0.228 & \textbf{37.10} & 71.6 & \textbf{1192} & \textbf{3113} \\
\midrule
\multirow{2}{*}{OTA2}
& w/o AP  & 55.59 & 1.249 & 80.27 & \textbf{78.2} & 3833 & \textbf{5873} \\
& APPlace & \textbf{59.83} & \textbf{1.734} & \textbf{88.86} & 77.1 & \textbf{3707} & 6263 \\
\midrule
\multirow{2}{*}{OTA3}
& w/o AP  & 55.27 & \textbf{1.456} & 68.30 & 79.9 & 4410 & 13234 \\
& APPlace & \textbf{59.13} & 1.235 & \textbf{81.26} & \textbf{83.9} & \textbf{4040} & \textbf{11864} \\
\bottomrule
\end{tabular}
}
\vspace{0.08cm}

\resizebox{0.85\columnwidth}{!}{
\begin{tabular}{clccccc}
\toprule
Design & Variant & Delay (ps) & Offset (mV) & E/cycle (fJ) & WL ($\mu$m) & Area ($\mu$m$^2$) \\
\midrule
\multirow{2}{*}{COMP}
& w/o AP  & 135.1 & 1.892 & 96.08 & 367 & 274 \\
& APPlace & \textbf{109.9} & \textbf{1.682} & \textbf{93.90} & \textbf{366} & \textbf{228} \\
\bottomrule
\end{tabular}
}
\vspace{-0.6cm}
\end{table}

With AP co-optimization, the post-layout results improve on all four circuits.
On the OTAs, gain increases by about 4~dB on every circuit, and CMRR, the metric most sensitive to layout imbalance and mismatch, improves by 1.4~dB to 13.0~dB, while routed wirelength decreases by 3.3\% to 8.4\%.
UGB, PM, and area change within small ranges and show no consistent trend.
On COMP, all five reported metrics improve, most notably the delay, which drops from 135.1~ps to 109.9~ps.
The two variants differ only in whether AP locations are co-optimized, so these gains are attributable to the joint optimization.

\subsubsection{Structure-preserving legalization}
\label{subsubsec:exp-ablation-lg}
{Hybrid} isolates the legalization stage, as it shares the same GP result with {APPlace} and differs only in the legalizer.
As reported in Tables~\ref{tab:ota_all_main} and~\ref{tab:comp_main}, replacing our legalizer degrades the post-layout results on every benchmark, most visibly the gain on OTA2 (59.83~dB to 41.55~dB) and the area on OTA3 (11864~$\mu$m$^2$ to 20769~$\mu$m$^2$).

\begin{figure}[tb]
    \centering
    \includegraphics[width=\linewidth]{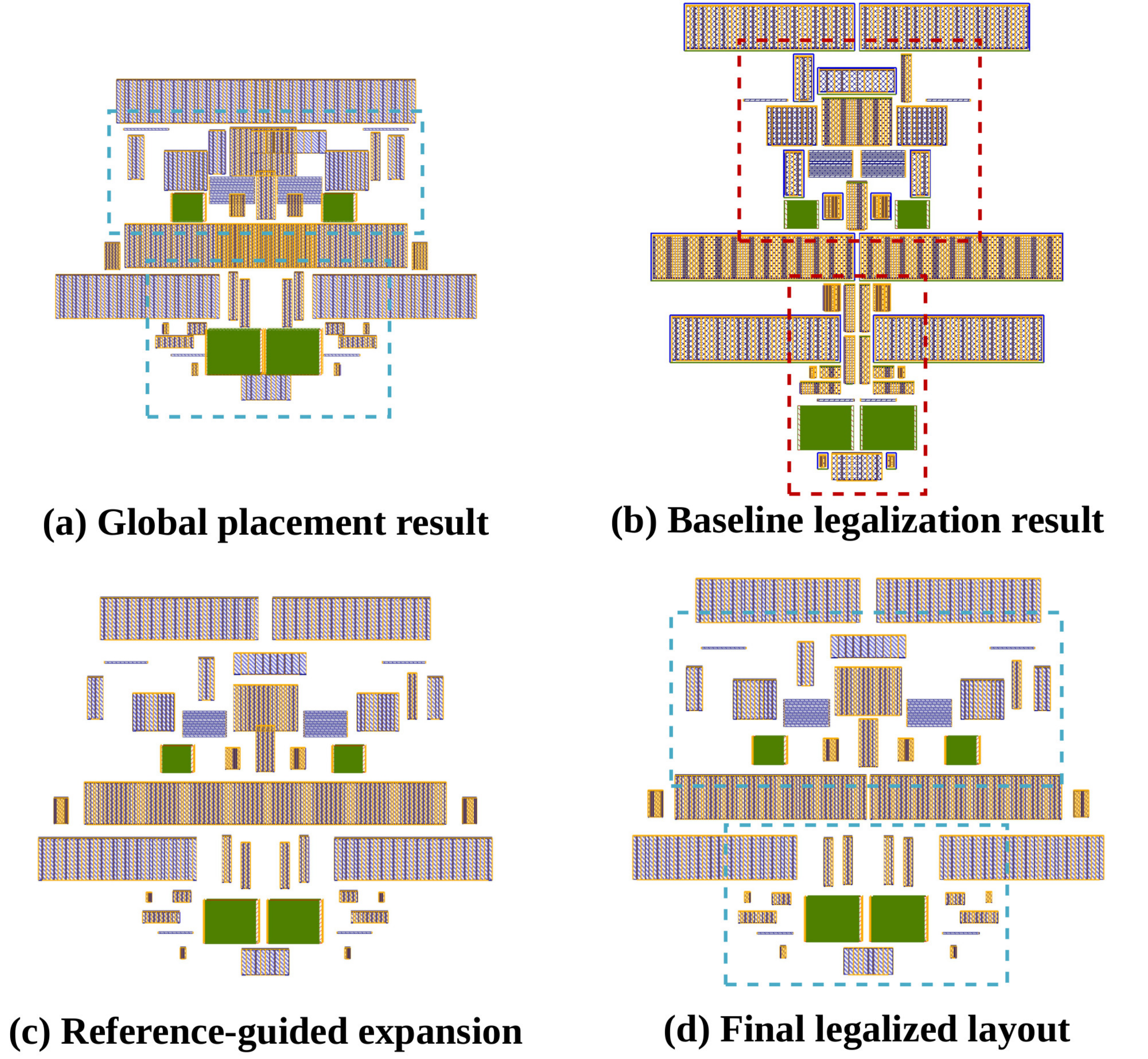}
    \Description{Four layout snapshots of OTA3 made of device rectangles. (a) The global placement result. (b) The baseline legalization result, with red dashed boxes marking regions whose arrangement is distorted relative to (a). (c) The layout after reference-guided expansion. (d) The final legalized layout from APPlace, with dashed boxes marking the same regions, whose arrangement follows (a).}
    \caption{
    Comparison of legalization trajectories on OTA3.
    The baseline~\cite{xu_magical_2019} flow is (a)--(b), while the APPlace flow is (a)--(c)--(d). The red dashed rectangles in (b) highlight the regions where the baseline legalizer severely distorted the topology, and the green rectangles in (d) show the same regions being preserved.
    }
    \label{fig:lg_compare}
    \vspace{-0.4cm}
\end{figure}

Figure~\ref{fig:lg_compare} shows the cause on OTA3.
Panel (a) is the routing-aware GP result before legalization.
Panel (b) is the layout produced by the MAGICAL legalizer, where large global movement visibly distorts the GP structure in the highlighted regions.
Panels (c) and (d) show the APPlace trajectory.
The reference-guided expansion first opens up the layout in a controlled manner, and the local resolution then removes the remaining conflicts while staying close to the GP geometry.
The preserved regions in panel (d) correspond to the distorted regions in panel (b).
These observations link the post-layout gap between the two flows to how well legalization preserves the structure established by GP.

\section{Conclusion}
\label{sec:conclusion}

We have presented APPlace, a routing-aware analog placement framework that treats device locations and AP selections as a joint optimization problem.
A differentiable superellipse AP geometry, an AP-aware mixed wirelength objective, and an overlap-gated adaptive control strategy form this joint optimization, and a structure-preserving legalization stage carries the jointly optimized structure through to routing.
Experimental results on real-world analog benchmarks show that APPlace improves the overall post-layout tradeoff over representative baselines, and the ablation studies attribute the gains to AP co-optimization and to structure preservation in legalization.
In future work, we plan to extend it to more diverse analog topologies and richer structural constraints.

\begin{acks}
This work is supported in part by NSF under grant CCF-2112665, Samsung, the UT Austin iMAGiNE consortium, and an equipment donation from NVIDIA.
\end{acks}

\balance
{
\scriptsize
\bibliographystyle{ACM-Reference-Format}
\bibliography{./ref/TOP_sim.bib, ./ref/DAC24.bib}
}

\end{document}